\documentclass{aa}
\usepackage[varg]{txfonts}
\usepackage{graphicx}

\begin{document}

\title{A semi-analytical light curve model and its application to type IIP supernovae}
\author{A. P. Nagy\inst{1}
\and A. Ordasi\inst{1}
\and J. Vinkó\inst{1,2} 
\and J. C. Wheeler\inst{2}}

\institute{Department of Optics and Quantum Electronics, University of Szeged, Hungary
\and Department of Astronomy, University of Texas at Austin, Austin, TX, USA}

\date{Accepted September 15, 2014}

\abstract{The aim of this work is to present a semi-analytical light curve model code which can be used for estimating physical properties of core collapse supernovae (SNe) in a quick and efficient way. To verify our code we fit light curves of Type II SNe and compare our best parameter estimates to those from hydrodynamical calculations. For this analysis we use the quasi-bolometric light curves of five different Type IIP supernovae. In each case we get appropriate results for the initial pre-supernova parameters. We conclude that this semi-analytical light curve model is useful to get approximate physical properties of Type II SNe without using time-consuming numerical hydrodynamic simulations.}

\maketitle 

\section{Introduction}
The light curves of Type IIP supernovae  are characterized by a plateau phase with a duration of about 80-120 days and a quasi-exponential tail caused by the radioactive decay of $^{56}$Co \citep[e.g][]{Maguire}. The emitted flux at later phases is directly determined by the mass of ejected $^{56}$Ni. Beside nickel mass other parameters specify the bolometric luminosity of SNe, such as explosion energy, ejected mass, and the initial size of the radiating surface \citep{Grassberg, Litvinova}. The radius of this surface is thought to be equal to the radius of the progenitor at the moment of shock breakout following core collapse. 
 
A general approach to determine the properties of supernova explosions is the  modeling of observed data with hydrodynamical codes \citep{Grassberg, Falk, Hillebrandt, Blinnikov,Nadyozhin, Utrobin07, Pumo1, Bersten11}. However, a simple analytical method may also be used to get approximate results \citep{Arnett80, Arnett82,Zampieri, Chatzopoulos}. With the help of these analytic light curve models, the basic physical parameters, such as the explosion energy, the ejected mass and the initial radius, can be estimated \citep{Arnett80, Popov}.  Although such simple estimates can be considered only preliminary, they can be obtained
without running complicated, time-consuming hydrodynamical simulations.  
Analytic codes may be useful in providing constraints for the most important physical parameters which can be used as input in more detailed simulations. Also, analytic codes may also give first-order approximations when the observational information is limited, for example when only photometry and no spectroscopy is available for a particular SN.   
    
In this paper we describe a semi-analytical light curve model which is based on the one originally developed by \cite{Arnett}. We assume a homologously expanding and spherically symmetric SN ejecta having a uniform density core and an exponential density profile in the other layers. Radiation transport is treated by the diffusion approximation. The effect of recombination causing the rapid change of the effective opacity in the envelope is taken into account in a simple form introduced by \citet{Arnett}.

This paper is organized as follows: in Section 2 we describe the model and its implementation, and also present the effect of variations of the initial input parameters on the calculated bolometric luminosity. In Section 3 we compare the results obtained for SN 2004et, SN 2005cs, SN 2009N, SN 2012A and SN 2012aw from our code and several hydrodynamic computations. Finally, Section 4 summarizes the main results of this paper.

\section{Main Assumptions and Model Parameters}
\subsection{The Light Curve Model}
We adopt the radiative diffusion model originally developed by \cite{Arnett80} and modified by \cite{Arnett} to take into account the effect of the recombination front in the ejecta. Below we review the original derivation, and present some corrections and implementations for numerical computations.

The first law of thermodynamics in Lagrangian coordinates for a spherical star may be written as
\begin{equation}
\frac{dE}{dt}+P \frac{dV}{dt}=\epsilon-\frac{\partial L}{\partial m}\ ,
\end{equation}
where $E$ is the internal energy per unit mass, $P$ is the pressure, $V = 1/\rho$ is the specific volume, $\epsilon$ is the entire energy production rate per unit mass and $L$ is the luminosity \citep{Arnett80, Arnett82}. In a radiation-dominated envelope the internal energy per unit mass is $E = a\ T^4 V$, and the pressure is $P = E/ 3 V$, where $a$ is the radiation density constant.
The energy loss is driven by photon diffusion, so we may use the following equation for the derivative of the luminosity
\begin{equation}
\frac{\partial L}{\partial m} = {1 \over {4 \pi r^2 \rho}} \frac{\partial L}{\partial r} 
= -\frac{a}{r^2 \rho}\ \frac{\partial }{\partial r}\left(\frac{c\ r^2}{3 \kappa \rho}\ \frac{\partial T^4}{\partial r} \right)\ ,
\end{equation}
where $\kappa$ is the mean opacity, $T$ is the temperature, and $\rho$ is the density.
Since the supernova ejecta expand homologously we define a comoving, dimensionless radius $x$ as 
\begin{equation}
r = R(t)\cdot x\ ,
\end{equation}
where $r$ is the distance of a particular layer from the center and $R(t)$ is the total radius of the expanding envelope at a given time. In the comoving coordinate system we are able to separate the time and space dependence of the physical properties. Thus, the density profile can be described as
\begin{equation}
\rho(x,t) = \rho(0,0)\ \eta(x) \left(\frac{R_0}{R(t)}\right)^3 ,
\end{equation}
where $R_0$ is the initial radius and $\eta(x) \sim \exp(-\alpha x)$ where $\alpha$ is assumed to be a small positive integer. The time-dependent term describes the dilution of the density due to expansion.

The ejecta are expected to be fully ionized shortly after the explosion, so it seems reasonable to consider a recombination front which moves inward through the envelope. The assumed recombination wave divides the ejecta into two different parts. The boundary separating the two layers occurs at the dimensionless radius $x_i$ where the local temperature $T(x)$ drops below the recombination temperature $T_{rec}$. Inside this recombination radius, the ejecta are assumed to be fully ionized. The opacity ($\kappa$) changes strongly at the boundary layer separating the two parts. Because the opacity has a strong non-linear dependence on the temperature we assume a simple step-function to approximate its behavior \citep{Arnett}:
\begin{equation}
\kappa(x,t)=\left
\{\begin{array}{rl}
\kappa_t& \textrm{, if }\ T(x)\geq T_{rec}\\
\\
0& \textrm{, if }\ T(x) < T_{rec}
\end{array}\right.
\end{equation}
where $T_{rec}$ is the recombination temperature, below which the ejecta are mostly neutral.
In this approximation we use the Thomson-scattering opacity for pure hydrogen gas as $\kappa \sim$ 0.4 $\mathrm{cm^2}$/g, and model the presence of heavier elements by setting $\kappa$ to lower values. For example $\kappa \sim$ 0.2 $\mathrm{cm^2}$/g is assumed for a He-dominated ejecta, while for a He-burned atmosphere $\kappa$ may be $\sim$ 0.1 $\mathrm{cm^2}$/g.   

\begin{figure}[!ht]
\includegraphics[width=9cm]{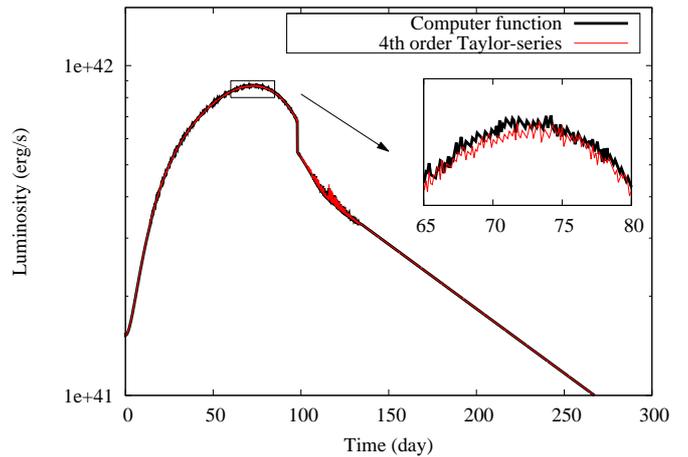}
\caption{Model light curve computed with exact $\sin(\pi x)/(\pi x)$ temperature profile (black) and with 4th order Taylor series (red).}
\label{fig1}
\end{figure}

Following \cite{Arnett80}, the temperature evolution and its spatial profile can be approximated as 
\begin{equation}
T^4(x,t) = T^4(0,0)\ \psi(x)\ \phi(t) \left(\frac{R_0}{R(t)}\right)^4 . 
\end{equation}
 and the radial components of this function is
\begin{equation}
\psi(x)\approx \frac{sin(\pi x)}{\pi x} ,
\end{equation}
which does not change during the expansion. While implementing the temperature profile in our C-code, we found that the direct application of the $\sin(x)$ function caused numerical instabilities due to rounding errors around $x \approx 1$. To reduce this problem, we used a 4th order Taylor-series expansion of the $\psi(x)$ function (see Fig.~\ref{fig1}). The implementation of the Taylor-series approximation increased the numerical stability when computing the effect of recombination.

Another important physical quantity is the energy production rate which can be defined as
\begin{equation}
\epsilon(x,t) = \epsilon(0,0)\ \xi(x)\ \zeta(t) .
\end{equation}
In this case a central energy production is assumed, which means that $\xi(x)$ is the Dirac-delta function at x = 0. The temporal dependence of the $\epsilon(t)$ function was specified by assuming either radioactive decay or magnetar-controlled energy input. In the following subsections we summarize these two different conditions.

\subsubsection{Radioactive energy input}
In this case only the radioactive decay of $^{56}$Ni and $^{56}$Co supplies the input energy. In such a model $\epsilon(0,0)$ is equal to the initial energy production rate of $^{56}$Ni-decay. The time-dependent part of the $\epsilon(x,t)$ function, when the ejecta are optically thick for gamma-rays, is given by
\begin{equation}
\zeta(t)= X_{Ni}+\frac{\epsilon_{Co}}{\epsilon_{Ni}} \ X_{Co} ,
\end{equation}
where $X_{Ni}$ and $X_{Co}$  are the number of nickel and cobalt atoms per unit mass, respectively, $\epsilon_{Ni}$ and $\epsilon_{Co}$ are the energy production rate from the decay of these elements. The number of the radioactive elements varies as
\begin{equation}
\frac{dX_{Ni}}{dz}= - X_{Ni}\quad \textrm{and}\quad \frac{dX_{Co}}{dz}=X_{Ni}-\mathbf{\frac{\tau_{Ni}}{\tau_{Co}}} X_{Co} ,
\end{equation}
where $z = t/\tau_{Ni}$ is the dimensionless time scale of the Ni-Co decay , $\tau_{Ni}$ and $\tau_{Co}$ are the decay time of nickel and cobalt, respectively.

The comoving coordinate of the recombination front is denoted as $x_i$. Following \cite{Arnett} it is assumed that the radiative diffusion takes place only within $x_i$ where $\kappa >$ 0, and the photons freely escape from $x = x_i$. Thus, the surface at $x_i$ acts as a pseudo-photosphere.

After inserting the quantities defined above into Eq. (1), we have
\begin{equation}
\frac{dE}{dt}+P \frac{dV}{dt} = \frac{a\ T^4(x,t)\ x_i^3\ V}{\phi(t)}\ \frac{d\phi(t)}{dt} + 2\ a\ T^4(x,t)\ x_i^2\ V\ \frac{dx_i}{dt} .
\end{equation} 

Now, applying the assumption made by \citet{Arnett82} (see his Eq. 13), the temporal and spatial parts of Eq. (11) 
can be separated, thus, both of them are equal to a constant (the ``eigenvalue'' of the solution). After separation,  the equation describing the temporal evolution of the $\phi(t)$ function can be expressed as
\begin{equation}
\frac{d\phi(t)}{dt}\ \tau_{Ni}=\frac{R(t)}{R_0\ x_i^3}\left[p_1 \zeta(t) - p_2\ x_i\ \phi(t) - 2\ \tau_{Ni}\ x_i^2\ \phi(t)\ \frac{R_0}{R(t)}\ \frac{dx_i}{dt}\right] ,
\end{equation}
where we corrected a misprint that occurred in Eq. (A41) of \cite{Arnett}. 
This equation contains two parameters defined as
\begin{equation}
p_1=\frac{\tau_{Ni} \epsilon_{Ni} M_{Ni}^0}{E_{Th}(0)}\quad \textrm{and}\quad p_2=\frac{\tau_{Ni}}{\tau_d} ,
\end{equation}
where $M_{Ni}^0=4 \pi \rho(0,0) R_0^3 \int\limits_0^{1} \xi(x) \eta(x) x^2 dx$ and $E_{Th}(0)=4 \pi R_0^3 a T^4(0,0) \int\limits_0^{1} \psi(x) x^2 dx$ are the initial total nickel mass and the initial thermal energy, respectively.

Eq. (12) was solved by the Runge-Kutta method with the approximation of 
$dx_i/dt \approx \Delta x_i/\Delta t$, where a small time-step of $\Delta t = 1$ s was applied.
In the n-th time-step $\Delta x_i = x_i^{(n)} - x_i^{(n-1)}$ was used, 
where $x_i^{(n)}$ and $x_i^{(n-1)}$ are the dimensionless
radii of the recombination layer in the $n$-th and $(n-1)$-th time-step, respectively.
To determine the value of $x_i^{(n)}$ in every time-step, our code divides the envelope into
thin ($\delta x = 10^{-9}$) layers, then calculates the temperature in each layer starting from
the outmost layer at $x = 1$ until the temperature exceeds the recombination temperature $T_{rec}$.
If that occurs at the $k$-th layer then $x_i \approx (x_k + \delta x/ 2)$ is chosen as the new radius of the recombination layer.

Finally, the total bolometric luminosity can be expressed as a sum of the radioactive heating plus the energy released by the recombination:  
\begin{equation}
L(t)= x_i\ \frac{\phi(t)\ E_{Th}(0)}{\tau_d} + 4\ \pi\ r_i^2\ Q\ \rho(x_i,t) \frac{dr_i}{dt} ,
\end{equation}
where  $\tau_d = 3\ \kappa\ \rho(0,0)\ R_0^2/(\pi^2 c)$ is the diffusion time scale, and $Q = 1.6\cdot 10^{13} (Z/A) Z^{4/3}$ is the recombination energy per unit mass. The effect of gamma-ray leakage can be taken into account as
\begin{equation}
L(t)= x_i\ \frac{\phi(t)\ E_{Th}(0)}{\tau_d}\ \left(1 - e^{-A_g/t^2} \right) + 4\ \pi\ r_i^2\ Q\ \rho(x_i,t)\ \frac{dr_i}{dt} ,
\end{equation}
where the $A_g$ factor refers to the effectiveness of gamma-ray trapping. The optical depth of gamma-rays can be defined as $\tau_g = A_g / t^2$ \citep{Chatzopoulos}. This parameter is significant in modeling the light curves of Type IIb and Ib/c SNe.

\subsubsection{Magnetar spin-down}
Magnetars represent a sub-group of neutron stars with a strong ($10^{14}$ - $10^{15}$G) magnetic field.
The spin-down power of a newly formed magnetar can creates a brighter and faster evolving SN light curve than radioactive decay does \citep{Piro}. This mechanism can contribute to the extreme peak luminosity of Type Ib/c and Super-Luminous SNe \citep{Woosley,Kasen, Chatzopoulos}.    

In this case, $\epsilon(t)$ includes radioactive energy production as well as magnetar spin-down:
\begin{equation}
\epsilon(t) = \epsilon_{Ni}(t) + \epsilon_{M}(t) ,
\end{equation}
where $\epsilon_{Ni}(t)$ is the energy production rate of radioactive decay of nickel and cobalt as defined in the previous section and $\epsilon_{M}(t)$ is the energy production rate of the spin-down per unit mass.
The power source of the magnetar is given by the spin-down formula
\begin{equation}
\epsilon_M(t) = \frac{E_p}{t_p\ M_{ej}}\ \frac{l-1}{(1 + t/t_p)^l} ,
\end{equation}
where $M_{ej}$ is the ejected mass, $E_p$ is the initial rotational energy of the magnetar, $t_p$ is the characteristic time scale of spin-down, which depends on the strength of the magnetic field, and $l = 2$ for a magnetic dipole.

Solving Eq.(1) in the same way as in the previous section the $\phi(t)$ function can be expressed as
\begin{eqnarray}
\frac{d\phi(t)}{dt}\ \tau_{Ni} & = & \frac{R(t)}{R_0\ x_i^3}\left[p_1 \zeta(t) 
- p_2\ x_i\ \phi(t)\ +\ p_3 \frac{1}{(1 + t/t_p)^2} \right ] - \nonumber \\
 & & - 2 \tau_{Ni}\ \phi(t)\ \frac{1}{x_i} \frac{dx_i}{dt}  \ , 
\end{eqnarray}
where $p_3 = \tau_{Ni} E_p/E_{Th}(0) t_p$. The total bolometric luminosity in this configuration is calculated using
Eq.(14)-(15) with the numerical integration of the modified $\phi(t)$ function.

To verify the magnetar model, we compared our results with the estimated peak luminosities defined by \cite{Kasen}:
\begin{equation}
L_{peak}^{ref} \approx \frac{E_p\ t_p}{t_d^2}\ \left[\ln\left(1+\frac{t_d}{t_p}\right)-\frac{t_d}{t_d + t_p}\right]\ ,
\end{equation}
where $t_d$ is the effective diffusion time scale \citep{Arnett80}:
\begin{equation}
t_d = \sqrt{\frac{2\ \kappa\ M_{ej}}{13.8\ v_{sc}\ c}}
\end{equation}
and $v_{sc} \approx \sqrt{10E_{SN}/3M_{ej}}$ is the characteristic ejecta velocity.

For the test case we used the following fixed parameters:
$R_0=5\cdot10^{11}$ cm; $M_{ej}=1 M_\odot$; $M_{Ni}^0=0 M_\odot$; $T_{rec}=0$ K (i.e. no recombination); $E_{kin}(0)=3$ foe ($1foe = 10^{51}$ erg); $E_{Th}(0)=2$ foe; $\alpha=0$ (constant density model);  $\kappa=0.34$ $\mathrm{cm^2/g}$; $A_g=10^6$ day$^2$ (full gamma-ray trapping). These parameters imply $v_{sc} \sim 22,400$ km s$^{-1}$ and $t_d \sim 14$ days. Values of $E_p$ and $t_p$ were varied within a range typical for magnetars \citep{Kasen}.
Table~\ref{table:1} shows the peak luminosities estimated from the formulae above ($L_{peak}^{ref}$) 
and provided by our code ($L_{peak}^{model}$). We found acceptable agreement between the calculated and the model values, although it is seen that the analytic formula slightly underestimates the model peaks.
 
\begin{table}[!ht]
\caption{Comparison of magnetar model peak luminosities with the analytic estimates}
\label{table:1}
\centering
\begin{tabular}{c c c c}
\hline
\hline \\
$E_p$ & $t_p$ & $L_{peak}^{ref}$  & $L_{peak}^{model}$  \\
($10^{51}$erg) & (day) & ($10^{44}$erg/s) & ($10^{44}$erg/s)\\
\hline  \\
1 & 5 & 1.76 & 1.94\\
5& 2  & 7.09 & 7.98 \\
5& 5 & 8.81 & 9.71 \\
5& 10 & 8.61 & 9.79\\
5& 50 & 4.15 & 5.89 \\
10 & 5  & 17.6 & 19.4 \\
\hline
\hline
\end{tabular}
\end{table}

\subsection{The Effect of Varying Input Parameters}
To create light curves we need to integrate Eq. (12)-(18) then apply Eq. (14)-(15) in every time-step. 
The input parameters for the model are the followings:
\begin{itemize}
\item[•]{$R_0$: Initial radius of the ejecta (in $10^{13}$ cm)}
\item[•]{$M_{ej}$: Ejected mass (in $\mathrm{M_\odot}$)}
\item[•]{$M_{Ni}^0$: Initial nickel mass (in $\mathrm{M_\odot}$)}
\item[•]{$T_{rec}$: Recombination temperature (in K)}
\item[•]{$E_{kin}(0)$: Initial kinetic energy (in foe)}
\item[•]{$E_{Th}(0)$: Initial thermal energy (in foe)}
\item[•]{$\alpha$: Density profile exponent }
\item[•]{$\kappa$: Opacity (in $\mathrm{cm^2/g}$)}
\item[•]{$E_p$: Initial  rotational energy of the magnetar (in foe)}
\item[•]{$t_p$: Time scale of magnetar spin-down (in day)}
\item[•]{$A_g$: Gamma-ray leakage exponent (in day$^2$)}
\end{itemize}

We tested our code by changing these input parameters and comparing the resulting light curves to those of \cite{Arnett}. The parameters were varied one by one using three different values while holding the others constant. 
The following reference parameters were chosen (plotted with black): $R_0=5\cdot10^{12}$ cm; $M_{ej}=10 \mathrm{M_\odot}$; $M_{Ni}^0=0.01 \mathrm{M_\odot}$; $T_{rec}=6000$ K; $E_{kin}(0)=1$ foe; $E_{Th}(0)=1$ foe; $\alpha=0$; $\kappa=0.3$ $\mathrm{cm^2/g}$; $E_p =0$ foe; $t_p=0$ days; $A_g=10^6$ day$^2$. When the magnetar energy input was taken into account the two characteristic parameters were $E_p =1$ foe and $t_p=10$ days. 
 
\begin{figure}[!ht]
\includegraphics[width=9cm]{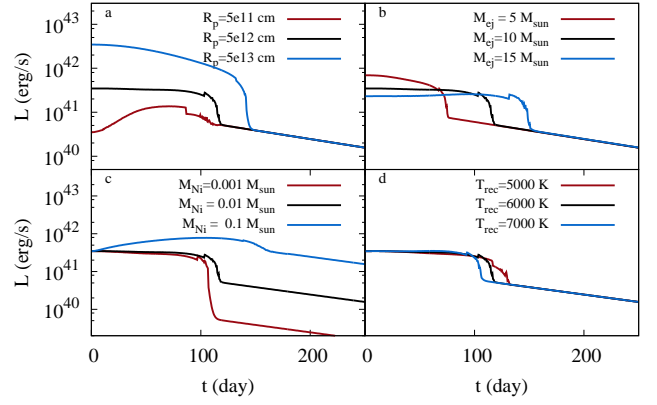}
\caption{Effect of changing the initial radius of the ejecta (panel \textbf{a}),
the ejected mass (panel \textbf{b}), the initial nickel mass (panel \textbf{c}) and 
the recombination temperature (panel \textbf{d}).}
\label{fig2}
\end{figure}

First, we created light curves with three different radii: $5\cdot10^{11}$, $5\cdot10^{12}$ and $5\cdot10^{13}$ cm. As it can be seen in Fig.~\ref{fig2}a, the modification of this parameter mainly influences the early part of the light curve. As a result of the increasing radius, the peak of the light curve becomes wider and flatter. The rapid decline after the plateau also becomes steeper. As Type IIP SNe exhibit steep declines after the plateau phase, this behavior is consistent with their larger progenitor radii. Our code  replicates this behavior.   
 
Next, we set 5, 10 and 15 $\mathrm{M_\odot}$ as the three input values of the ejected mass. This parameter also affects the maximum and the width of the light curve (Fig.~\ref{fig2}b). Higher ejecta masses result in lower peak luminosities and more extended plateau phases. 

The three different values of nickel mass were 0.001, 0.01 and 0.1 $\mathrm{M_\odot}$. Fig.~\ref{fig2}c shows the strong influence of this parameter on all the phases of the light curve. Increasing nickel mass causes a global increase of the luminosity at all phases, as expected.  This panel also illustrates that the late-phase luminosity level depends on only the Ni-mass in
the case of full gamma-ray trapping.

Fig.~\ref{fig2}d shows the effect of the modification of the recombination temperature from 5000 K to 7000 K. This parameter has no major influence on the light curve. Higher $T_{rec}$ results in a shorter plateau phase and the short decline phase after the plateau also becomes steeper. The recombination temperature is the parameter that can be used to take into account the ejecta chemical composition. For example, the recombination temperature for pure H ejecta is $\sim$ 5000 K, but for He-dominated ejecta it is higher,  $T_{rec} \sim 7000$ K \citep{GraNad76}, or $T_{rec} \sim 10,000$ K \citep{Hatano99} .    

\begin{figure}[!ht]
\includegraphics[width=9cm]{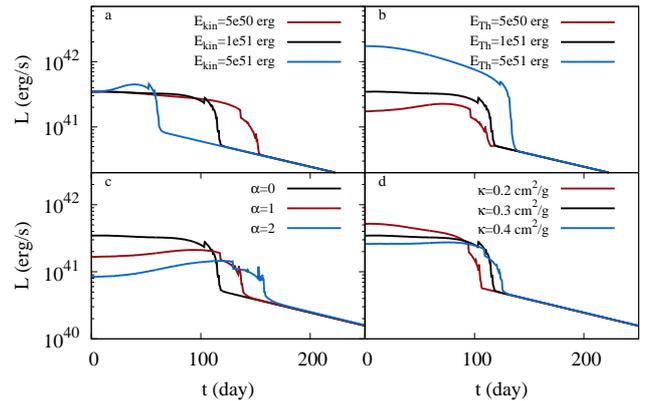}
\caption{Effect of changing the initial kinetic energy (panel \textbf{a}),
the initial thermal energy (panel \textbf{b}), the density profile exponent (panel \textbf{c})
and the opacity (panel \textbf{d}).}
\label{fig3}
\end{figure}

One of the most important parameters of SN events is the explosion energy ($E_{SN}$) which is the sum of the kinetic and the thermal energy. In this work we examined the effect of these two parameters separately.    
The three values of the kinetic energy were 0.5, 1 and 5 foe. As Fig. \ref{fig3}a shows, this parameter has significant influence on the shape of the early light curve but does not have any effect on the late part because, again, the luminosity at late phases are set by only the initial Ni-mass. When the kinetic energy is lower, the plateau becomes wider, while the maximum luminosity decreases. Note that using extremely high kinetic energy results in the lack of the plateau phase.
The influence of the thermal energy is somewhat similar: it affects mainly the early light curve (Fig.~\ref{fig3}b). Increasing thermal energy widens the plateau, and the peak luminosity rises. For high $E_{Th}$ the plateau phase starts to disappear, just as for high $E_{kin}$.

The density profile exponent was chosen as 0, 1 and 2. Fig.~\ref{fig3}c. shows that the different values cause changes mainly in the maximum of the light curve and the duration of the plateau phase. 
If the density profile exponent is higher, then the luminosity is reduced and the peak becomes wider.   

Next, we examined the effect of changing opacity (Fig.~\ref{fig3}d). The chosen values of this parameter were 0.2, 0.3 and 0.4 $\mathrm{cm^2 g^{-1}}$. Decreasing the opacity results in rising luminosity and shorter plateau phase.

\newpage

\begin{figure}[!ht]
\includegraphics[width=9cm]{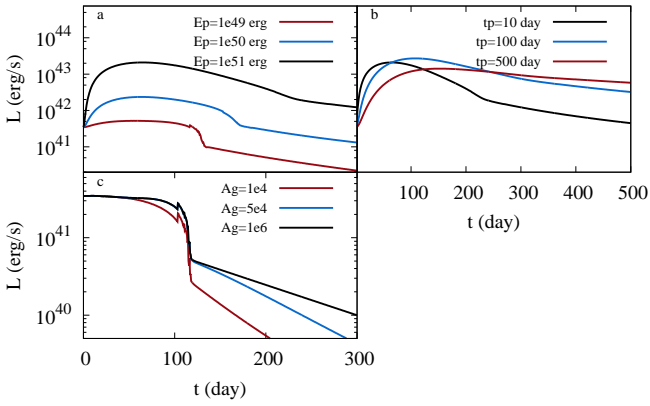}
\caption{Effect of changing the initial rotation energy of the magnetar (panel \textbf{a}), 
the characteristic time scale of magnetar spin-down (panel \textbf{b}), the gamma-ray leakage exponent (panel \textbf{c}).}
\label{fig4}
\end{figure}

To test the magnetar energy input, the initial rotational energy of the magnetar were set as 0.01, 0.1 and 1 foe. As Fig.~\ref{fig4}a shows, this parameter has significant influence on the entire light curve. Higher $E_p$ results in rising luminosities and broader plateau phase. If the initial rotational energy is not comparable to the recombination energy, no recognizable plateau phase is created by the magnetar energy input.

The characteristic time scale of the spin-down was chosen as $t_p = 10$, 100 and 500 days. The light curve strongly depends on the ratio of the effective diffusion time and spin-down timescale (Fig.~\ref{fig4}b). As far as $t_p$ is well below $t_d$, increasing spin-down time causes higher luminosities and wider plateau phase, but if $t_p >> t_d$, the maximum starts to decrease. In this particular case $t_d \sim 97.35$ days was applied.   

Finally, the gamma-ray leakage exponent was varied as $10^4$, $5\cdot10^4$ and $10^6\ \mathrm{day^2}$. Fig.~\ref{fig4}c shows the strong influence of this parameter on the entire light curve.The tail luminosity is significantly related to the gamma-ray leakage, as expected. Increasing $A_g$ results in a wider plateau phase, and also increases the tail luminosity.

To summarize the results of these tests, we conclude that
\begin{itemize}
\item[a)]{the duration of the plateau phase is strongly influenced by the values of the initial radius of the ejecta, the ejected mass, the density profile exponent and the kinetic energy;}  
\item[b)]{the opacity, the initial nickel mass and the recombination temperature are weakly correlated with the duration of the plateau;} 
\item[c)]{the maximum brightness and the form of the peak mainly depend on the thermal energy, the initial nickel mass, the initial radius of the ejecta, the density profile exponent and the magnetar input parameters;}
\item[d)]{the peak luminosity of the plateau is weakly influenced by the ejected mass and the opacity; }
\item[e)]{the late light curve is determined by the amount of the initial nickel mass, the gamma-ray leakage exponent and the characteristic features of the magnetar;}
\item[f)]{the light curve is less sensitive to the recombination temperature and opacity.}
\end{itemize}
These results are generally in very good agreement with the conclusions by \cite{Arnett} regarding the behavior of the initial radius, the recombination temperature and the factor $\kappa\ M_{ej}/v_{sc} \approx \kappa\ M_{ej}/\sqrt{E_{SN}/M_{ej}}$. Furthermore our results show the same parameter dependence of the plateau duration as calculated by \cite{Popov}.

\subsection{Parameter correlations}
\begin{figure}[!ht]
\includegraphics[width=9cm]{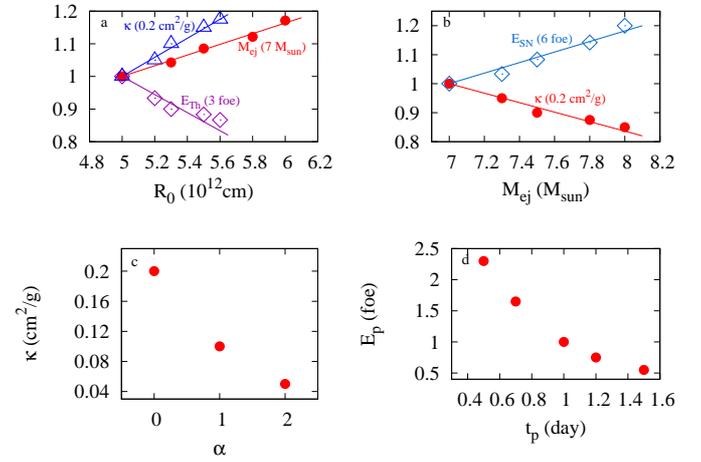}
\caption{Scatter diagrams of the correlated parameters. Panel \textbf{a}: $\kappa$ (in $\mathrm{0.2 cm^2/g}$), $M_{ej}$ (in $\mathrm{7 M_\odot}$) and $E_{Th}$ (in 3 foe) vs. $R_0$; 
panel \textbf{b}: $E_{SN}$ (in 6 foe) and $\kappa$ (in $\mathrm{0.2 cm^2/g}$) vs. $M_{ej}$; panel \textbf{c}: $\kappa$ vs. $\alpha$; Panel \textbf{d}: $E_p$ vs. $t_p$ }
\label{fig5}
\end{figure}

The correlation between parameters were examined by the Pearson correlation coefficient method which measures  the linear correlation between two variables. For this comparison, we first synthesized a test light curve for    
both the radioactive decay and magnetar-controlled energy input models. Then we tried each parameter-combination to create the same light curve and determine the correlations among the parameters. The scatter diagrams (Fig.~\ref{fig5}) illustrate this correlation between the two particular parameters: if the general shape of the distribution of random parameter choices shows a trend, the parameters are more correlated. 
 
The final result shown in Fig.~\ref{fig5} suggests that only three of the parameters are independent: $T_{rec}$, $M_{Ni}^0$ and $A_g$ while the other parameters are more-or-less correlated with each other.  

\section{Comparison with observations and hydrodynamic models}

In this section we compare the parameters calculated from our radioactive energy input models with those from hydrodynamic simulations. We fit SNe 2004et, 2005cs, 2009N, 2012A and 2012aw by our code using the radioactive energy input. Since
our simple code is unable to capture the first post-breakout peak, the comparison between the data and the model was
restricted to the later phases of the plateau and the radioactive tail.
 
Like most other SN modeling codes, our code needs the bolometric light curve, which is not observed directly. In order to assemble the bolometric light curves for our sample we applied the following steps. First, the measured magnitudes in all available photometric bands were converted into fluxes using proper zero points \citep{Bessell}, extinctions and distances. The values of extinction in each case were taken from the NASA/IPAC Extragalactic Database (NED). At epochs when an observation with a certain filter  was not available we linearly interpolated the flux using nearby data. For the integration over wavelength we applied the trapezoidal rule in each band with the assumption that the flux reaches zero at 2000 $\mathrm{\dot{A}}$. The infrared contribution was taken into account by the exact integration of the Rayleigh-Jeans tail from the wavelength of the last available photometric band (I or K) to infinity.

Note that to get a proper comparison with the other models collected from literature, we calculated the bolometric light 
curve using the same distance as in the reference papers. 

\subsection{SN 2004et}
SN 2004et was discovered on 2004 September 27 by S. Moretti \citep{Zwitter}. It exploded in a nearby starburst galaxy NGC 6946 at a distance of about 5.9 Mpc. This was a very luminous and well-observed Type IIP supernova \citep{Sahu} in optical (UBVRI) and NIR (JHK) wavelengths. In this paper all of these photometric bands were used to derive the bolometric light curve. 

\begin{figure}[!ht]
\includegraphics[width=9cm]{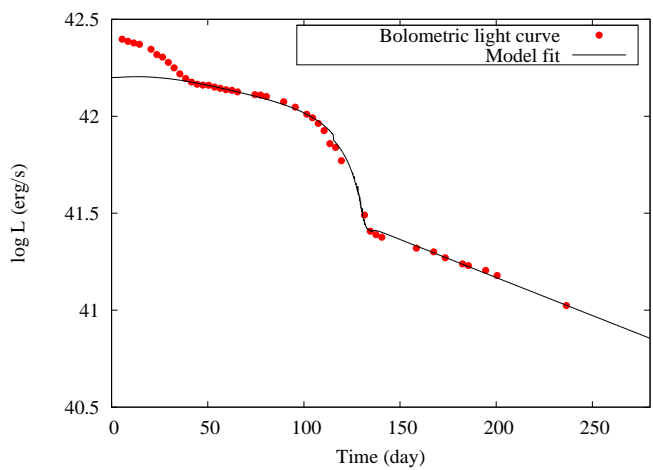}
\caption{Light curve of SN 2004et (dots) and the best result of our model (solid line).}
\label{fig6}
\end{figure}

 In the literature SN 2004et was modeled with different approaches. \cite{Utrobin09} used a 1-dimensional hydrocode to estimate the progenitor mass and other physical properties. \cite{Maguire} applied the formulae by \cite{Litvinova} that are based on their
hydrodynamical models, and also used the steepness parameter method from \cite{Elmhamdi}, to get the physical parameters of SN 2004et.
Table~\ref{table:2} shows the parameters from \cite{Maguire} and \cite{Utrobin09} as well as our best-fit results. 
The bolometric light curve of SN 2004et and the model curve fitted by our code are plotted in Fig.~\ref{fig6}.   

\begin{table}[!ht]
\caption{Results for SN 2004et} 
\label{table:2} 
\centering                  
\begin{tabular}{l c c c}          
\hline
\hline \\                      
Parameter & This paper &\multicolumn{2}{c}{Literature}\\
& & Model A\tablefootmark{1} & Model B\tablefootmark{2} \\ 
\hline  \\                         
$R_0$ [$10^{13}$ cm]& 4.2 & 4.39 & 10.4\\
$M_{ej}$ [$\mathrm{M_\odot}$]& 11.0 & 14.0 & 22.9\\
$M_{Ni}$ [$\mathrm{M_\odot}$]& 0.060 & 0.060 & 0.068\\
$E_{tot}$ [$10^{51}$erg]& 1.95 & 0.88 & 2.30\\
\hline 
\hline                                            
\end{tabular}
\tablebib{(1)~\citet{Maguire}; (2)~\citet{Utrobin09}.}
\end{table}

\subsection{SN 2005cs}
\begin{figure}[!ht]
\includegraphics[width=9cm]{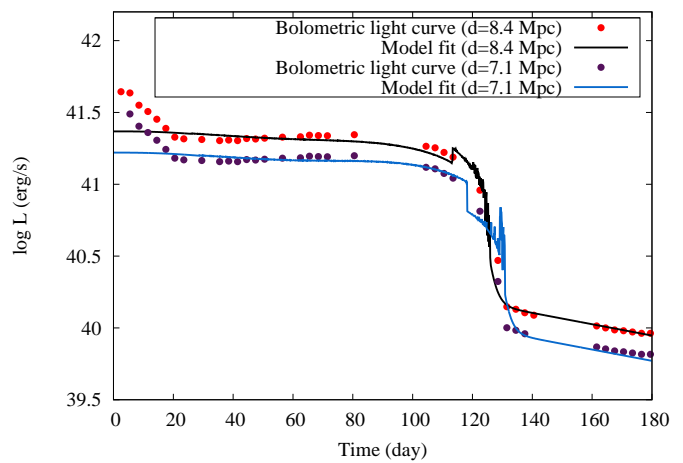}
\caption{Best fit for SN 2005cs (black line) and the bolometric luminosity at 8.4 Mpc (red dots). The orange circles represent the light curve of SN 2005cs at 7.1 Mpc and the gray line is our fit.}
\label{fig7}
\end{figure}

The under-luminous supernova SN 2005cs was discovered on 2005 June 30 in M51 by \cite{Kloehr}. This event was more than a magnitude fainter than an average Type IIP supernova. Nevertheless the light cure of SN 2005cs was observed in UBVRI bands and its physical properties were calculated by \cite{Tsvetkov} based on the hydrodynamic model of \cite{Litvinova}. SN 2005cs was also fitted by our code and the results can be found in Table~\ref{table:3}. For better comparison we used d = 8.4 Mpc for the distance of M51 which was adopted by \cite{Tsvetkov}. However, we also calculated the quasi-bolometric light curve with d = 7.1 Mpc \citep{Takats1}. The results for both distances are listed in Table~\ref{table:3}. The best fit of our model can be seen in Fig.~\ref{fig7}.     
\begin{table}[!h]
\caption{Results for SN 2005cs} 
\label{table:3}     
\centering                  
\begin{tabular}{l c c c}          
\hline
\hline \\                      
Parameter &  \multicolumn{2}{c}{This paper} & Literature\tablefootmark{1}\\ 
& d = 7.1 Mpc & d = 8.4 Mpc \\
\hline  \\                         
$R_0$ [$10^{13}$ cm]& 1.20 & 1.50 & 1.22\\
$M_{ej}$ [$\mathrm{M_\odot}$]& 8.00 & 8.00 & 8.61\\
$M_{Ni}$ [$\mathrm{M_\odot}$]& 0.002 & 0.003 & 0.0018\\
$E_{tot}$ [$10^{51}$erg]& 0.48 & 0.5 & 0.3\\
\hline 
\hline                                            
\end{tabular}
\tablebib{(1)~\citet{Tsvetkov}.}
\end{table}

\subsection{SN 2009N}
\begin{figure}[!ht]
\includegraphics[width=9cm]{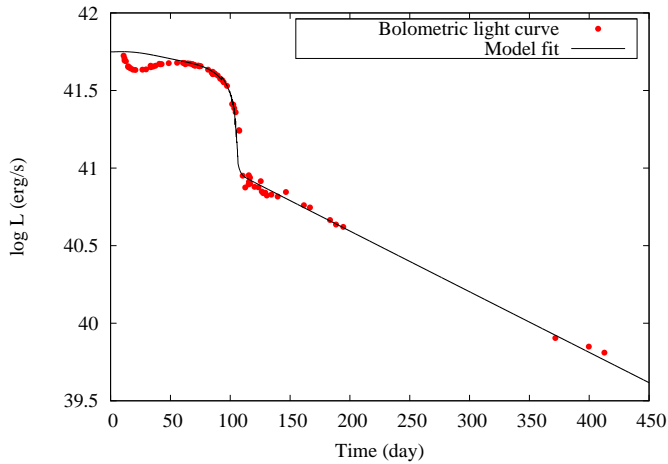}
\caption{Solid line shows the fit of our model and the dots represent bolometric luminosity from observed data of SN 2009N.}
\label{fig8}
\end{figure}

SN 2009N was discovered in NGC 4487 having a distance of 21.6 Mpc \citep{Takats2}. The first images were taken by Itagaki on 2009 Jan. 24.86 and 25.62 UT \citep{Nakano}. This event was not as luminous as a normal Type IIP SN, but it was brighter than the under-luminous SN 2005cs. Hydrodynamic modeling was presented by \cite{Takats2} who applied the code of \cite{Pumo1} and \cite{Pumo2}. In Table~\ref{table:4} we summarize our results as well as the properties from hydrodynamic simulations. Fig.~\ref{fig8} shows the luminosity from the observed data and the model light curve.
\begin{table}[!h]
\caption{Results for SN 2009N}
\label{table:4}
\centering                  
\begin{tabular}{l c c}          
\hline
\hline \\                      
Parameter & This paper & Literature\tablefootmark{1}\\ 
\hline  \\                         
$R_0$ [$10^{13}$ cm]& 1.60 & 2.00 \\
$M_{ej}$ [$\mathrm{M_\odot}$]& 7.6 & 11.5 \\
$M_{Ni}$ [$\mathrm{M_\odot}$]& 0.016 & 0.02\\
$E_{tot}$ [$10^{51}$erg]& 0.8 & 0.48 \\
\hline 
\hline                                            
\end{tabular}
\tablebib{(1)~\citet{Takats2}.}
\end{table}

\subsection{SN 2012A}
\begin{figure}[!ht]
\includegraphics[width=9cm]{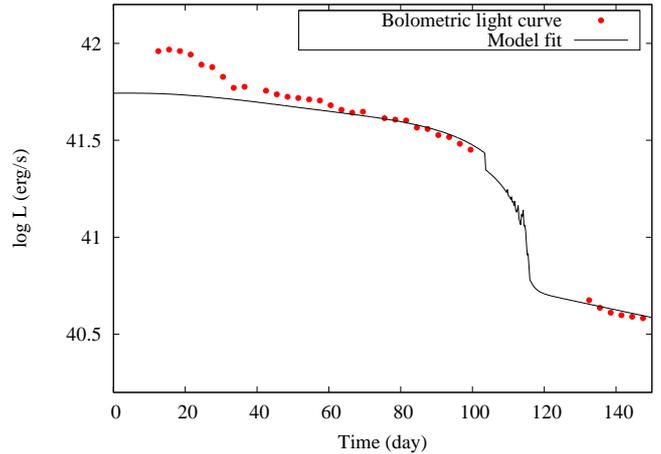}
\caption{Light curve of SN 2012A (dots) and the best result of our model (solid line).}
\label{fig9}
\end{figure}

SN 2012A was discovered in an irregular galaxy NGC 3239 at a distance of 9.8 Mpc \citep{Tomasella}. The first image was taken on 2012 Jan 7.39 UT by \cite{Moore}. This event was classified as a normal Type IIP supernova with a short plateau. The luminosity drop after the plateau was intermediate between those of normal and under-luminous Type IIP SNe. The fact that SN 2012A exploded in a nearby galaxy made this object very well observed in multiple (UBVRIJHK) bands. For computing the physical properties of the progenitor, \cite{Tomasella} applied a semi-analytical \citep{Zampieri} and a radiation-hydrodynamic code \citep{Pumo1, Pumo2}. Table~\ref{table:5} contains the final results of \cite{Tomasella} and our fitting parameters as well. Our model light curve can be seen in Fig.~\ref{fig9}.
\begin{table}[!h]
\caption{Results for SN 2012A}  
\label{table:5}
\centering                  
\begin{tabular}{l c c}          
\hline
\hline \\                      
Quantity & This paper & Literature\tablefootmark{1}\\ 
\hline  \\                         
$R_0$ [$10^{13}$ cm]& 1.8 & 1.8 \\
$M_{ej}$ [$\mathrm{M_\odot}$]& 8.80 & 12.5 \\
$M_{Ni}$ [$\mathrm{M_\odot}$]& 0.01 & 0.011 \\
$E_{tot}$ [$10^{51}$erg]& 0.8 & 0.48\\
\hline 
\hline                                            
\end{tabular}
\tablebib{(1)~\citet{Tomasella}.}
\end{table}

\subsection{SN 2012aw}
SN 2012aw was discovered on 2012 March 16.86 UT by P. Fagotti \citep{Fagotti} in a spiral galaxy M95 at an average distance of 10.21 Mpc. This was a very well-observed Type IIP supernova in optical (UBVRI) and NIR (JHK) wavelengths \citep{Dallora}. All of these photometric bands were used to create the bolometric light curve.

\begin{figure}[!ht]
\includegraphics[width=9cm]{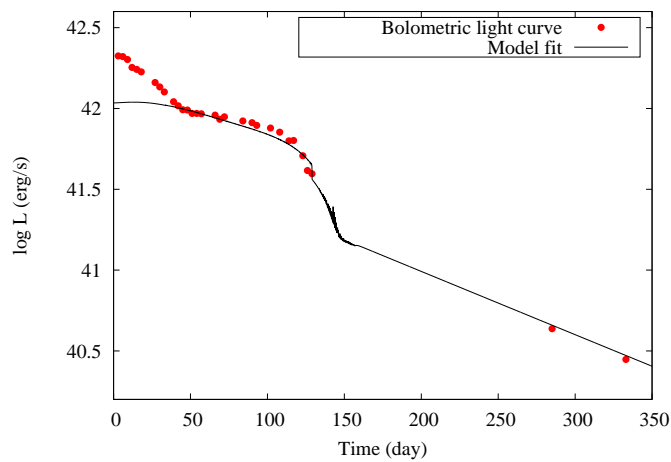}
\caption{Light curve of SN 2012aw (dots) and the best result of our model (solid line).}
\label{fig10}
\end{figure}

The physical properties of SN 2012aw modeled by \cite{Dallora} who applied two different codes: a semi-analytic \citep{Zampieri} and a radiation-hydrodynamic code \citep{Pumo1, Pumo2}. In  Table~\ref{table:6} we summarize our result as well as the parameter values of \cite{Dallora}. Our best-fit model is plotted in Fig.~\ref{fig10}.

\begin{table}[!h]
\caption{Results for SN 2012aw}
\label{table:6}      
\centering                  
\begin{tabular}{l c c}          
\hline
\hline \\                      
Quantity & This paper & Literature\tablefootmark{1}\\ 
\hline  \\                         
$R_0$ [$10^{13}$ cm]& 2.95 & 3.0 \\
$M_{ej}$ [$\mathrm{M_\odot}$]& 20.0 & 20.0 \\
$M_{Ni}$ [$\mathrm{M_\odot}$]& 0.056 & 0.056 \\
$E_{tot}$ [$10^{51}$erg]& 2.2 & 1.5\\
\hline 
\hline                                            
\end{tabular}
\tablebib{(1)~\citet{Dallora}.}
\end{table}

\section{Discussion and Conclusion}
The good agreement between the results from the analytical light curve modeling and the parameters from other hydrodynamic calculations leads to the conclusion that the usage of the simple analytical code may be useful for preliminary studies prior to more expensive hydrodynamic computation. The code described in this paper is also capable of providing quick estimates for the most important parameters of SNe such as the explosion energy, the ejected mass, the nickel mass and the initial radius of the progenitor from the shape and the peak of the light curve as well as its late-phase behavior. 
Note that there is a growing number of observational evidence showing that the plateau durations have a narrow distribution 
with a center at about 100 days \citep{Faran}, which suggests that strong correlations exist between parameters like the explosion energy and the progenitor radius in the presupernova stage. The code may offer a fast and efficient way to explore such kind
of parameter correlations.

Tables 2-6 show that the hydrodynamic models for Type IIP events consistently give higher ejected masses than our code. 
On the other hand, there are also major differences between the values given by different hydrocodes, as e.g. in the case of 
SN~2004et. Although the total SN energies from our code are usually higher then those obtained from more complex models, 
they show a similar trend: for an under-luminous SN the best-fit energy is lower, while for a more luminous SN the code suggests higher explosion energy. 

The present code has various limitations and caveats. One of them is that it is not able to fit the light curve at very early epochs. This may be explained by the failure of the assumption of homologous expansion at such early phases, and/or the adopted simple form of the density profile.  Another possibility can be the assumption of a two-component ejecta configuration which is sometimes used for modeling Type IIb SNe \citep{Bersten}. In this case the model also contains a low-mass envelope on top of the inner, more massive core.  Within this context the fast initial decline may be due to radiation diffusion from the 
fast-cooling outer envelope heated by the shock wave due to the SN explosion. These models will be studied in detail in a forthcoming paper.

\begin{acknowledgements}
This work has been supported by Hungarian OTKA Grant NN 107637 (PI Vinko) and NSF Grant AST 11-09801 (PI Wheeler). 
We express our sincere thanks to an anonymous referee for the very thorough report that helped us a lot while
correcting and improving the previous version of this paper.  
\end{acknowledgements}

\end{document}